# The early Universe with JWST and ALMA

Rodrigo Herrera-Camus[1,2]*, Natascha Förster Schreiber[3], Livia Vallini[4], Rychard Bouwens[5] & John D. Silverman[6,7,8]

*Corresponding author: Rodrigo Herrera-Camus (rhc@udec.cl)

The Atacama Large Millimeter/submillimeter Array and the James Webb Space Telescope are transforming our understanding of galaxy formation and evolution in the early Universe. By combining their capabilities, these observatories provide unprecedented insights into the gas, dust, and stars of high-redshift galaxies at spatially resolved scales, unveiling the complexities of their interstellar medium, kinematics, morphology, active galactic nuclei, and star formation activity. This review summarizes recent breakthroughs in the study of galaxies during the first billion years of cosmic history, highlighting key discoveries, open questions, and current limitations. We discuss how observations, theoretical models, and simulations are shaping our understanding of early galaxy evolution and identify promising directions for future research. While significant progress can be achieved through optimized use of existing facilities and collaborative efforts, further advances will require enhanced angular resolution and sensitivity, motivating upgrades to current instruments and the development of next-generation observatories.

Over the past three years, Atacama Large Millimeter/submillimeter Array (ALMA) and the James Webb Space Telescope (JWST) have provided an unparalleled multi-wavelength view of the gas, dust, and stars in galaxies at kiloparsec scales, extending our reach from the local Universe to the first galaxies. The field of galaxy evolution has entered a golden era: a decade ago, multi-wavelength studies of multi-phase gas, dust, and stars at kiloparsec scales were limited to nearby galaxies. Today, this capability extends to the first galaxies in the Universe at redshift $z\approx10$ and beyond. These observations now allow us to address fundamental questions by directly probing the earliest stages of galaxy formation and evolution: How did galaxies look like in the first billion years? What were their physical conditions? How fast and through which pathways were heavy elements, dust, and central supermassive black holes (SMBHs) produced? What was the role of gas outflows driven by feedback from supernovae, massive stars radiation and winds, and active galactic nuclei (AGN) powered by accreting SMBHs in regulating star formation activity and chemical enrichment?

In December 2024, an international group of researchers working on galaxy evolution gathered at the Lorentz Center Workshop *Synergistic ALMA+JWST View of the Early Universe* to examine recent progress enabled by the combination of these facilities. This review summarizes key discoveries and ways forward discussed in three main areas: (1) the interstellar medium (ISM), (2) galaxy kinematics and structure, and (3) outflows and active galactic nuclei (AGN).

## Interstellar Medium

Figure 1 illustrates how the synergy between ALMA and JWST allows for a comprehensive study of all major components of the ISM in star-forming galaxies at $z\gtrsim4$. These include the radiation field produced by stars, ionized gas in HII regions, the interface with neutral gas in photodissociation regions (PDRs), molecular gas, as well as interstellar dust, including polycyclic aromatic hydrocarbons (PAHs), and metals (elements heavier than He).

Dust profoundly impacts the physics, chemistry, and observed properties of galaxies. Among the foremost ALMA discoveries, substantial dust reservoirs have been unexpectedly detected in systems as distant as $z\approx8$, when the universe was merely 600 million years old[1,2,3]. This finding challenges existing dust formation theories, calling for a revision of mechanisms such as dust production in supernova ejecta and grain growth in the ISM[4,5]. With JWST, measurements of ultraviolet/optical attenuation curves in galaxies at $z\approx7$, including the 2175 Å

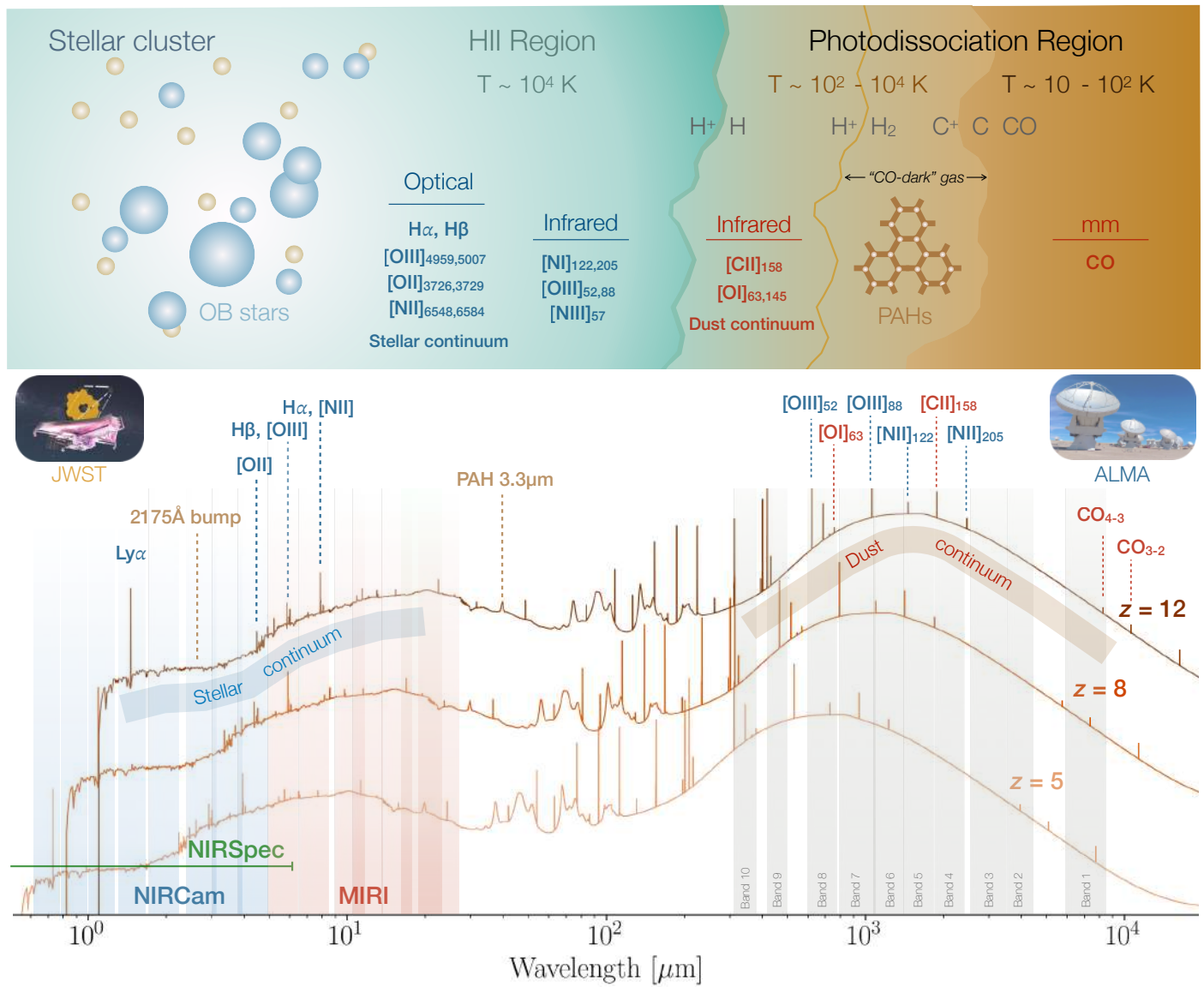

**Fig. 1 | ALMA and JWST views of stars and the ISM in high-redshift galaxies.** Top: schematic representation of the interplay between stars, ionized gas in H II regions, neutral gas in PDRs, molecular clouds and dust—including PAHs—along with the key spectral tracers of these components across the rest-frame UV, optical, infrared and millimetre wavelengths. Bottom: SED of a typical star-forming galaxy in the local Universe, modelled with CIGALE[98] and redshifted to $z = 5$, 8 and 12. The lefthand side highlights the coverage provided by JWST/NIRCam and MIRI broadband filters, and the spectrograph NIRSpec, enabling studies of the stellar continuum, nebular emission lines from H II regions, the 2,175-Å PAH extinction feature and the 3.3-μm emission feature for $z \approx 4$–6 galaxies. The right-hand side shows the spectral coverage of ALMA bands 1–10, which probe dust continuum emission, far-infrared cooling lines from various ISM phases, and CO transitions tracing molecular gas. Credit: JWST illustration, NASA GSFC/CIL/

bump[6] traditionally associated with more evolved galaxies, highlights the rapid emergence of carbonaceous grains and underscores the efficiency of early dust production processes. The direct detection of the 3.3 μm PAH feature in a dusty star-forming galaxy at $z \approx 4$[7] reinforces this picture.

JWST and ALMA are shedding new light on the related issue of how and when the first heavy elements enriched galaxies in the early universe. The detection of CIII] and CIV UV emission lines in galaxies just 350 million years after the Big Bang[8,9] ($z \approx 12$) suggests that carbon may have formed much earlier than previously thought, potentially making it one of the first metals to emerge in the universe, consistent with observations of carbon-enhanced old metal-poor stars in our Galaxy[10]. Detection of [O III] 88 μm emission in the galaxy JADES-GS-z14-0[11,12] at $z \approx 14$, supported by photometric measurements of optical [O III] and Hβ lines, indicates that this system is surprisingly metal-enriched ($Z \approx 0.05$– 0.2 $Z_\odot$) only 300 Myr after the Big Bang. These bright galaxies at z≳10 show very low dust attenuation, a phenomenon that cannot be explained solely by dust destruction but may instead result from outflows displacing dust to larger distances[13].

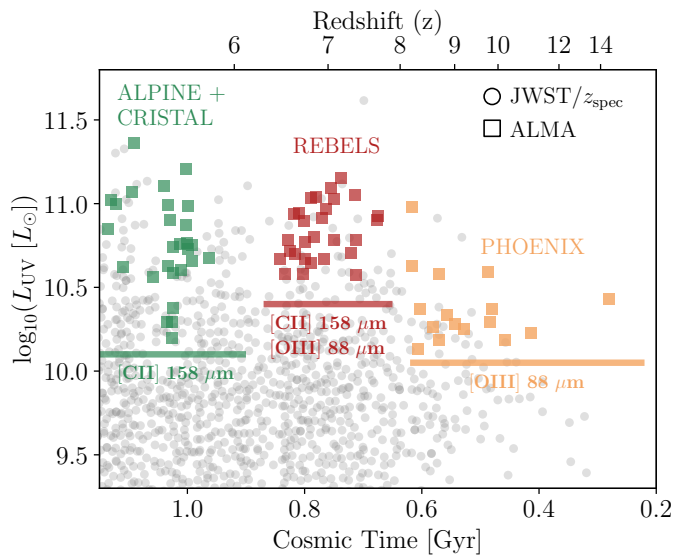

**Fig. 2 | High-redshift galaxies observed with ALMA and JWST.** UV luminosity (LUV) as a function of cosmic time (or redshift) for spectroscopically confirmed galaxies observed with JWST and with ALMA. The ALMA sample includes galaxies targeted by the Large Programs ALPINE[21–23] and CRISTAL[24] (green), REBELS[25] (red) and the recently awarded PHOENIX (orange; principal investigator S. Schouws), whose observations are scheduled for ALMA cycle 12. The main far-infrared transitions used in each study are also indicated.

With JWST/NIRSpec, derivation of metallicities up to z≈10 via the 'direct-$T_{e'}$' method using faint auroral emission lines now enables new calibrations for strong-line diagnostics, important for future large spectroscopic surveys[14,15]. However, important challenges remain: joint analyses combining optical ([O II]λλ3726,3729 and [O III]λ4363) and far-infrared ([O III] 88 μm and [O III] 52 μm) lines show that the direct method may underestimate metallicities by up to 0.8 dex, due to low-density gas not fully traced by optical lines alone[16]. Recent JWST data at z≈4–9 further show that electron densities in early galaxies are significantly higher than at z≈0, following approximately $n_e \propto (1+z)^{1-2}$, consistent with high ionization parameters and intense star-formation surface densities observed at high redshift[17]. Unexpectedly high N/O ratios found in some galaxies at z ≳ 6 indicate rapid enrichment and possibly unusual stellar populations[18,19]. The origin of this nitrogen excess remains uncertain but may provide key insights into early star formation and feedback[20].

Among the key gas tracers accessible with ALMA are the [C II] 158 μm transition, a primary coolant of the cold neutral ISM, and the [O III] 88 μm transition. As Fig. 2 shows, over the past five years, large ALMA programs — such as ALPINE[21,22,23] and CRISTAL[24] at z≈4-6, and REBELS[25] at z≈6-8— made headway in measuring these far-infrared emission lines in massive star-forming galaxies lying on the stellar mass vs. star formation rate "main sequence". These observations show that typical galaxies at z≳6 generally exhibit high [O III]/[C II] luminosity ratios (≳5)[26,27,28], with only a few cases displaying comparable [C II] and [O III] luminosities[29]. These findings support models in which intense starburst episodes drive bright [O III] emission. Cosmological zoom-in simulations suggest that bursty star formation may be a key factor in shaping the population of massive, bright galaxies identified by JWST[30,31], although this remains debated[32].

Important challenges remain in constraining dust and metal properties in high-redshift galaxies. Most dust estimates rely on a single-band continuum measurement, leaving large uncertainties in dust temperature ($T_{dust}$), infrared luminosity, and emissivity. Theoretical models that combine dust continuum and [CII] emission can help break the dust mass ($M_{dust}$) - dust temperature ($T_{dust}$) degeneracy[33]. Accurate dust constraints are also critical for resolving the age-dust degeneracy in modeling the Spectral Energy Distribution (SED) and thus for tracing galaxy assembly[35,35]. ALMA's high-frequency bands (8–10), which sample the dust SED peak, are essential for reducing $T_{dust}$ uncertainties[36,37] and clarifying obscured star formation in the early Universe.

Major ongoing efforts are devoted to understanding how the different phases of the ISM evolve with redshift — in particular their density, metallicity, and irradiation from star formation and AGN activity. Cosmological zoom-in simulations, such as the SERRA suite[38,39] shown in Fig. 3, can now predict key line ratios on resolved sub-galactic scales (down to ~10 pc) both in the rest-frame optical and UV[40,41] and rest-frame far-infrared[38,42,] to guide the derivation of the ISM properties. The census of rest-optical nebular lines is rapidly growing with the multiplexing and sensitivity of JWST/NIRSpec but the far-infrared regime remains largely unexplored. ALMA can access nearly all major far-infrared lines at high-redshifts (see Fig. 1; e.g., [O III] 52 μm, [N III] 57 μm, [O I] 63 μm, [N II] 122 μm, [O I] 145 μm, [C II] 158 μm, and [N II] 205 μm), though multi-line data have so far been limited to luminous or lensed galaxies[43]. Combining far-infrared with UV/optical tracers provides novel insights, such as using CIII]/[C II] to assess starburst activity[44], or [O III] optical and far-infrared line ratios to refine metallicity estimates by removing the dependence with gas temperature[45]. Expanding ALMA observations to bright z≳8 systems discovered by JWST (like the recently awarded ALMA Large Program PHOENIX; PI Schouws; see Fig. 2), or ALMA multi-line surveys of star-forming galaxies at z≳4, will boost our knowledge of the high-redshift ISM.

Tracing the molecular gas reservoir in high-redshift galaxies, and understanding its efficiency in fueling star formation, remains difficult at high-redshift. Low metallicities enhance CO photodissociation, leading to faint CO lines. Moreover, the ALMA, and NOEMA interferometers primarily probe higher-J CO lines at z≳4, which trace dense, warm gas rather than the total molecular reservoir. The [C II] 158 μm transition offers an alternative tracer[46,47], but its multi-phase origin and dependence on local ISM conditions[47] and star formation activity burstiness[48] complicate its interpretation. However, significant theoretical progress has been made[39], and observational advances are possible with current capabilities: a census of the gas content in high-$z$ galaxies can be obtained by combining dynamical masses from deep ALMA [C II] observations with robust JWST/NIRCam stellar mass estimates. Complementary [N II] 205 μm data can help account for the ionized [C II] component, while HI constraints from [C II] and [O I] observations[49,50], or from damping-wing absorption of the Lyman-α transition[51], will be crucial for a complete view of the gas reservoir.

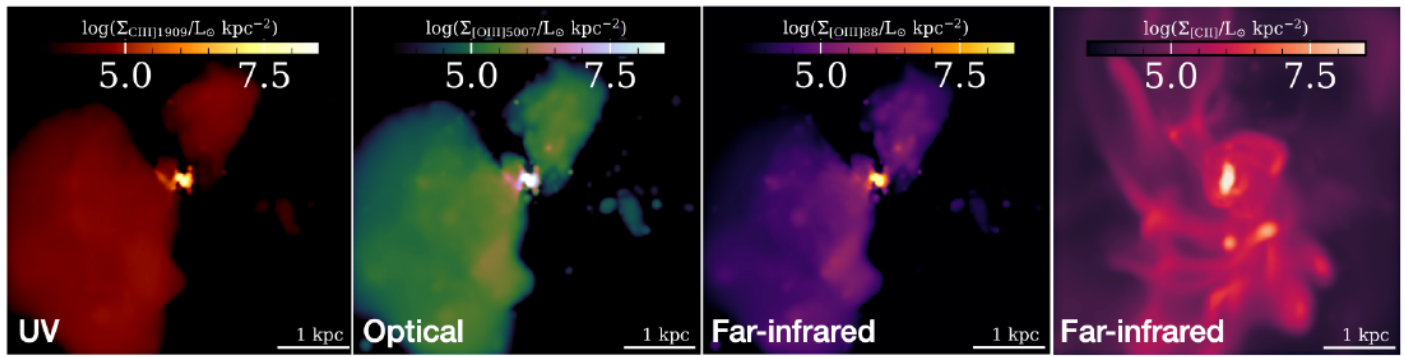

**Fig. 3 | Cosmological zoom-in simulation of a star-forming galaxy at $z = 6.5$.** Emission properties of a typical star-forming galaxy at $z=6.5$ from the SERRA simulations[38,39]. From left to right, the panels show the surface brightness of major UV, optical and far-infrared transitions: C III] 1,909 Å, [O III] 5,007 Å, [OIII] 88 μm and [CII] 158 μm.

In the near future, the ALMA Wideband Sensitivity Upgrade[52] is expected to enhance continuum and spectral line imaging speeds by a factor of ~2-3. This improvement will facilitate multi-band surveys of high-redshift star-forming galaxies, enabling more robust characterization of their dust SEDs and ISM conditions. Increased sensitivity will help to follow-up on the large population of galaxies uncovered by JWST at $z\gtrsim5$, many of which are too faint to be detectable with current ALMA sensitivities. As illustrated in Fig. 2, ALMA follow-up so far has focused on the brightest or most massive systems at each redshift. Looking further ahead, increasing ALMA's sensitivity by an additional factor of 5-10 will be critical for pushing into the regime of (unlensed) low-stellar-mass galaxies at $z\gtrsim4$, as well as the earliest galaxies that JWST will continue to discover at $z\gtrsim10$.

## Kinematics and Structure

Observations of the structure and kinematics of high-$z$ galaxies are essential for understanding the processes that drive galaxy assembly and disk formation. ALMA and JWST now indicate that galaxy disks may already be in place very early in the Universe[53,54,55], aligning with recent cosmological zoom-in simulations that predict the emergence of cold disks at $z > 6$ (ref. 56). Among massive ($\gtrsim 10^9$ M$_\odot$), typical star-forming galaxies from the CRISTAL survey at $4 \lesssim z \lesssim 6$, roughly 50% can be classified as disks. About half of them show no clear signs of interactions, while the other half exhibit features of rotating disks with some irregularities that could be induced by internal gravitational instabilities and/or nearby lower-mass companions[57]. This result is consistent with the fraction of observed disks at $z\approx5$ based solely on JWST/NIRCam imaging data[58]. At earlier cosmic times, two notable recent examples of rotating disks at $z\approx6$ and $z\approx7$ are the Cosmic Grapes[59] and REBELS-25[60], respectively, for which deep, high-resolution ALMA [CII] observations revealed high ratios of ordered-to-random motions.

Notably, despite their clumpy, irregular rest-UV/optical appearance in JWST/NIRCam imaging, some $z\gtrsim4$ galaxies exhibit regular disk-like [CII] kinematics, echoing earlier findings at $z\approx1-3$ from HST imaging and Hα or CO line kinematics[61]. Fig. 4 showcases the star-forming disk CRISTAL-08[24,57] at $z=4.43$, which hosts at least eight ~kilo-parsec size star-forming clumps visible in UV and optical light . The Cosmic Grapes is another such striking example, where gravitational lensing resolves the galaxy into at least 15 star-forming clumps, each roughly 10–60 pc in size[58] plausibly formed via gravitational instabilities (Toomre Q ~ 0.2-0.3) in its gas-rich disk with high surface mass densities comparable to local dusty starbursts. An interesting case is the submillimeter galaxy GN20 at $z=4.05$, for which deep JWST/NIRSpec Hα observations reveal clumpy and asymmetric emission, and large-scale disk kinematics yet with deviations from circular motions potentially indicative of radial inflows fueling bulge growth and AGN activity[62], akin to several cases reported at lower redshift[63].

Extending these pioneering kinematic studies of typical $z\gtrsim4$ galaxies to larger samples will be key to map out the early evolution of the disk fraction, the relative distribution of baryonic and dark matter, and the importance of non-circular motions, and to catch the first galactic disks to form. Current challenges will need to be overcome: the relatively small number of spectroscopically-confirmed galaxies especially at $z\gtrsim8$, their compactness and faintness. The highest spectral resolution of JWST/NIRSpec ($R=2700$) limits detailed kinematic studies and high-angular-resolution ALMA follow-up observations remain challenging given its current sensitivity (see Fig. 2). Crucially, high signal-to-noise ratio (S/N) is paramount to accurately determine dark matter fractions from extended rotation curves, and high angular and velocity resolution in the inner regions is essential to robustly constrain the presence of dense bulge components and signatures of radial gas motions. Future slit and integral field unit spectroscopy with the MICADO and HARMONI first-generation instruments on the Extremely Large Telescope (ELT), respectively, and the planned upgrades in ALMA sensitivity, will boost progress in this area.

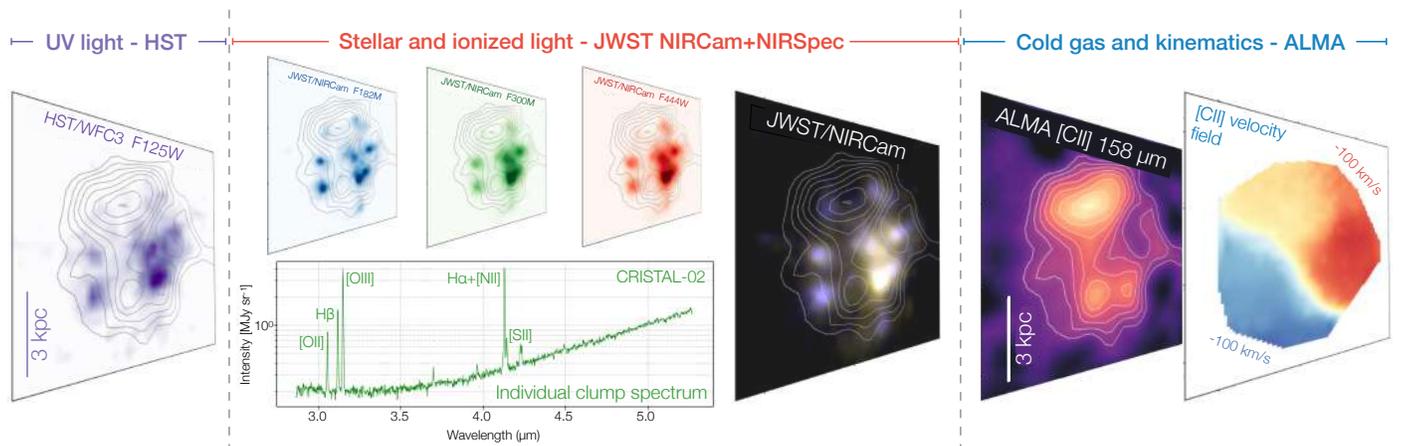

**Fig. 4 | Multiwavelength view of a star-forming galaxy at z ≈ 4.5 with ALMA, HST and JWST.** Left to right: for CRISTAL-08, a rest-frame UV emission map from young stars observed with HST; a stellar light map combining JWST/NIRCam filters F182M, F300M and F444W and a JWST/NIRSpec spectrum of a giant stellar clump from CRISTAL-02, another star-forming galaxy at z = 5.3[99,100]; an ALMA map of the cold gas traced by the integrated [C II] line emission, along with the corresponding velocity field, which—despite the clumpy morphology—exhibits signatures of smooth, ordered rotation. The contours show the integrated [C II] line emission.

The detection of very extended [CII] 158 µm line emission around high-redshift galaxies, often referred to as "[CII] halos"[64,65] has drawn considerable attention. In main-sequence, star-forming galaxies at z≈4-6 from the CRISTAL survey, the [C II] line emission reaches radii on average 2.9 and 1.5 times larger than the stellar rest-frame UV and dust far-infrared continuum, respectively[66]. The origins of this extended emission are likely diverse, involving unresolved galaxy companions, outflows expelling gas beyond the disk, inflowing gas, or gravitational interactions. Theoretical models[67] and refined zoom-in simulations that better resolve small-scale processes in the circumgalactic medium[68] are testing the physical mechanisms responsible for these halos.

Understanding the nature of [CII] halos requires a comprehensive observational approach: JWST/NIRCam imaging to identify potential minor companions, deep JWST/NIRSpec and ALMA spatially resolved observations to detect AGN activity and characterize extended gas in multiple phases, and Lyα spectroscopy to provide additional constraints from line profile differences. The CRISTAL-01 system at z=4.55 illustrates the power of the ALMA–JWST synergy: extended [CII] and Lyα emission appears partially driven by an outflow from AGN activity in the neighboring dusty star-forming galaxy J1000+0234, with multiple minor companions present, including one irradiated by the AGN that resembles the nearby Hanny's Voorwerp object[69,70].

## Outflows and AGNs

One of the most remarkable findings from the initial years of JWST operations is the unexpectedly high abundance of black holes in the early Universe[71,72]. These include black holes with large masses ($M_{BH} = 10^6$–$10^8$ $M_\odot$) identified at early cosmic times (z≈4-11). As Fig. 5 shows, many of them appear excessively massive compared to their host galaxies (up to $M_{BH}/M_\star$~0.01-0.1), strongly deviating from the local $M_{BH} - M_\star$ relation[73]. While further exploration is required, since the stellar masses of these systems might be underestimated, the black hole mass measurements could carry significant uncertainties, and selection effects play a role[74,75,76], it is clear that these findings challenge existing theoretical models of black hole formation in the early Universe[77,78].

In the spotlight are the so-called "little red dots" (LRDs)[79,80] uncovered at z≳4 by JWST: point-like, extremely red objects with highly unusual UV-optical colors. Initially suggested to be dusty starbursts, recent ALMA and NOEMA non-detections of far-infrared continuum emission disfavor this scenario[81]. An alternative is that LRDs host Type I AGN, with the BLR and accretion disk visible but attenuated by dust, or embedded in dense gas producing a non-stellar Balmer break[82]. Whether LRDs are compact massive galaxies, AGN, or both remains unclear[83,84]. A recent breakthrough came with the first direct dynamical black hole mass measurement in a strongly lensed LRD at z=7.04, which revealed a Keplerian rotation curve consistent with a ~5×10^7 $M_\odot$ black hole and little room for any stellar host component. This result supports the interpretation of at least some LRDs as 'naked' black hole seeds in their earliest growth phase[85].

Ways forward will focus on reducing uncertainties in measuring both black hole and stellar masses. If super-Eddington accretion dominates over standard models[86], black hole masses could be up to an order of magnitude lower than estimates based on local scaling relations[87]. For stellar masses, a major challenge is disentangling the host galaxy from the AGN contribution in SED modeling

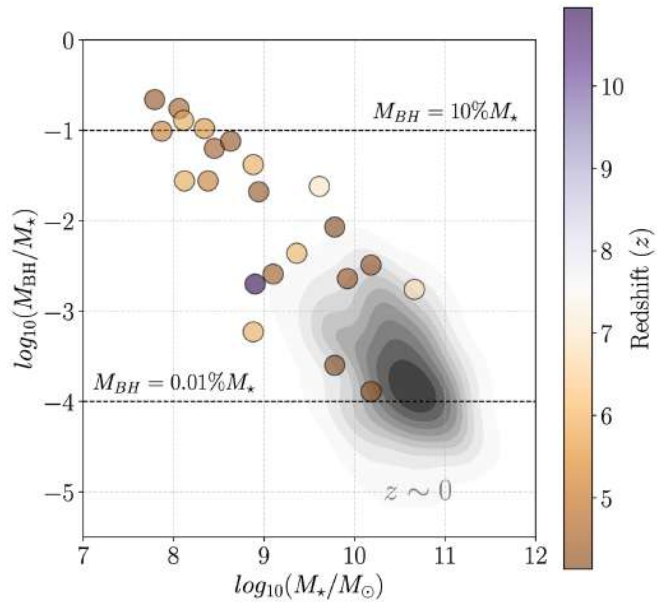

**Fig. 5 | Black hole–stellar mass relation in nearby and $z \gtrsim 4$ AGN galaxies.** The ratio of black hole mass to stellar mass in high-redshift AGN galaxies detected with JWST[71,72] is shown as a function of stellar mass, colour coded according to redshift. For comparison, the local AGN population[73] is shown as a grey cloud. Unlike the case for nearby AGNs, high-redshift black holes tend to be much more massive compared with their host galaxies, up to $M_{BH}/M_\star \approx 0.01$–$0.1$.

—studies have shown that neglecting or improperly modeling the AGN light can bias $M_\star$ estimates in AGN-dominated systems[88]. Additionally, high-redshift AGN appear to be significantly offset from the AGN locus in rest-optical nebular line ratios observed in the local Universe[71,72,89], making it difficult to use traditional diagnostic diagrams to distinguish AGN activity from star formation in these early systems. These limitations underscore the need for more refined models and observational techniques to properly assess the incidence and properties of black holes at high redshifts. Observationally, future synergies with GRAVITY+ and ELT/MICADO and HARMONI will provide robust mass estimates from direct BLR kinematics[90], and ≲100-pc resolution observations mapping the distribution and motions of gas and stars into the sphere of influence of distant massive black holes. These advances will also aid in the search for intermediate mass black holes and help constrain black hole seeding and growth pathways[91].

The surprising abundance and properties of $z \gtrsim 4$ SMBHs re-shapes the question on their role in regulating star formation and enriching the circumgalactic medium in metals in the early Universe alongside feedback from intense star formation. A hallmark of both AGN and star formation feedback is gas outflows, with JWST spectroscopy rapidly building a census of outflows across the population of high redshift systems. For instance, the incidence fraction of ionised outflows, traced by broad velocity components in emission line profiles, is approximately 25–40% in low-mass ($10^{7-9}\ M_\odot$) galaxies at $z \gtrsim 4$ (ref. 92). These ionized gas outflows exhibit typical velocities of 350 km s$^{-1}$ and mass outflow rates implying mass loading factors of $\eta \approx 2$, highlighting their efficiency in expelling gas relative to the depletion rate by star formation.

In nearby galaxies, the cold gas phase dominates the mass budget of ejected gas[93]. At high redshift, however, tracing this component in emission remains difficult. A powerful probe of molecular outflows is OH absorption[94], though this method is limited to bright, often lensed continuum sources at high redshift. A promising alternative is [C II] line emission, but distinguishing its molecular and atomic components in outflows is currently not possible. Studies combining observations of local outflows with simulations offer a valuable way forward for disentangling the relative roles of stellar clumps and AGN in powering feedback. Nevertheless, these comparisons must be carried out with caution, as the ISM state and structure at high redshift differ substantially from the local Universe and strongly influence how feedback couples to the gas.

ALMA observations of [C II] have revealed the cold phase of outflows in massive star-forming galaxies at $z \gtrsim 4$, through line stacking in large surveys such as ALPINE[95] and CRISTAL[96], as well as deep observations of individual systems[48,97]. A striking case is CRISTAL-02 at $z \approx 5$, where a giant star-forming clump drives a powerful outflow detected in [CII], and in H$\alpha$ and [O III] with JWST/NIRSpec. The inferred mass outflow rate of $\approx 500\ M_\odot$ yr$^{-1}$ implies that, without fresh gas accretion, CRISTAL-02 could quench within only 100 Myr (Davies et al., submitted manuscript).

Ideally, outflow studies at high redshift should trace both warm (e.g., H$\alpha$, [O III]) and cold (e.g., [C II], OH) phases, and use sensitive, spatially resolved observations to constrain size, inclination, and opening angle. Imaging is also crucial to rule out confusion with tidal features from interactions. Achieving this goal requires the synergy of ALMA and JWST.

## Summary

The combined capabilities of ALMA and JWST are revolutionizing our understanding of galaxies in the early Universe. These observatories provide spatially and spectroscopically resolved, multi-wavelength views of early galaxies, allowing detailed measurements of their gas, dust, and stellar content, as well as insights into their kinematics, morphology, and star formation up to unprecedented redshifts. While significant progress has been made, many key questions remain, as highlighted in this review. Some can be addressed with current facilities through coordinated efforts focused on deeper, higher angular resolution observations and more systematic studies of diverse galaxy populations. Others will require upcoming capabilities of on-going developments or entirely new advancements. At long wavelengths, much needed sensitivity gains are on the horizon notably with the ongoing ALMA Wideband Sensitivity Upgrade and potential further fivefold or greater increase in the longer term. At short wavelengths, diffraction-limited instruments such as MICADO and HARMONI at the ELT will achieve six times higher angular resolution than JWST at comparable sensitivity, opening up unique opportunities to resolve stellar populations, ionized gas, and star-forming structures in galaxies at $z>4$, complementing ALMA's access to the cold ISM. With these advancements, the future of galaxy formation and evolution studies is exceptionally promising.

## Affiliations

1 Departamento de Astronomía, Universidad de Concepción, Concepción, Chile. 2 Millennium Nucleus for Galaxies, Concepción, Chile. 3 Max-Planck-Institut für extraterrestrische Physik, Garching, Germany. 4 INAF-Osservatorio di Astrofisica e Scienza dello Spazio, Bologna, Italy. 5 Leiden Observatory, Leiden University, Leiden, The Netherlands. 6 Kavli Institute for the Physics and Mathematics of the Universe (WPI), The University of Tokyo Institutes for Advanced Study, The University of Tokyo, Kashiwa, Japan. 7 Center for Data-Driven Discovery, Kavli IPMU (WPI), UTIAS, The University of Tokyo, Kashiwa, Japan. 8 Department of Astronomy, School of Science, The University of Tokyo, Tokyo, Japan. 9 Center for Astrophysical Sciences, Department of Physics & Astronomy, Johns Hopkins University, Baltimore, MD, USA.

## Author Contributions

R.H.-C., N.M.F.S., L.V., R.B., and J.D.S. organized the Lorentz Workshop *Synergistic ALMA+JWST View of the Early Universe*, which led to this review article and provided the basis for its content. They wrote the text; R.H.-C. made the figures; and L.V. provided the SERRA simulations shown in Fig. 3. All authors participated in the discussion of the results and contributed to revising the manuscript.

## Acknowledgments

We kindly thank the Lorentz Center for hosting the *Synergistic ALMA+JWST View of the Early Universe* workshop, where we had the opportunity to discuss the progress and main results that ALMA and JWST have produced in the study of the early Universe, which serve as the basis for this review article. We would also like to thank all the workshop participants, whose contributions during the small-group and plenary discussions form the foundation of this article. We are grateful to the anonymous referees for their insightful and constructive comments, which significantly improved the clarity and overall quality of the manuscript. We thank S. Schouws for kindly providing the data shown in Figure 2. R.H.-C. thanks the Max Planck Society for support under the Partner Group project "The Baryon Cycle in Galaxies" between the Max Planck Institute for Extraterrestrial Physics and the Universidad de Concepción. R.H.-C. also gratefully acknowledges financial support from ANID–MILENIO–NCN2024112 and ANID BASAL FB210003. N.M.F.S. acknowledges funding by the European Union (ERC Advanced Grant GALPHYS, 101055023). Views and opinions expressed are, however, those of the authors only and do not necessarily reflect those of the European Union of the European Research Council. Neither the European Union nor the granting authority can be held responsible for them. LV acknowledges support from the INAF Minigrant "RISE: Resolving the ISM and Star formation in the Epoch of Reionization" (Ob. Fu. 1.05.24.07.01).

## References

1. Tamura, Y. et al. Detection of the Far-infrared [O III] and Dust Emission in a Galaxy at Redshift 8.312: Early Metal Enrichment in the Heart of the Reionization Era. *Astrophys. J.* **874**, 27 (2019).
2. Witstok, J. et al. Dual constraints with ALMA: new [O III] 88-μm and dust-continuum observations reveal the ISM conditions of luminous LBGs at z ~ 7. *Mon. Not. R. Astron. Soc.* **515**, 1751–1773 (2022).
3. Algera, H. S. B. et al. The ALMA REBELS survey: the dust-obscured cosmic star formation rate density at redshift 7. *Mon. Not. R. Astron. Soc.* **518**, 6142–6161 (2023).
4. Mauerhofer, V. & Dayal, P. The dust enrichment of early galaxies in the JWST and ALMA era. *Mon. Not. R. Astron. Soc.* **526**, 2196–2210 (2023).
5. Palla, M. et al. Metal and dust evolution in ALMA REBELS galaxies: insights for future JWST observations. *Mon. Not. R. Astron. Soc.* **528**, 2407–2423 (2024).
6. Witstok, J. et al. Carbonaceous dust grains seen in the first billion years of cosmic time. *Nature* **621**, 267–270 (2023).
7. Spilker, J. S. et al. Spatial variations in aromatic hydrocarbon emission in a dust-rich galaxy. *Nature* **618**, 708–712 (2023).
8. D'Eugenio, F. et al. JADES: Carbon enrichment 350 Myr after the Big Bang. *Astron. Astrophys.* **689**, 152 (2024)
9. Castellano, M. et al. JWST NIRSpec Spectroscopy of the Remarkable Bright Galaxy GHZ2/GLASS-z12 at Redshift 12.34. *Astrophys. J.* **972**, 2 (2024).
10. Placco, V. et al. Carbon-enhanced Metal-poor Star Frequencies in the Galaxy: Corrections for the Effect of Evolutionary Status on Carbon Abundances. *Astrophys. J.*, **797**, 1, (2014).
11. Carniani, S. et al. The eventful life of a luminous galaxy at z=14: metal enrichment, feedback, and low gas fraction. *Astron. Astrophys.* **696**, 87 (2025)
12. Schouws, S. et al. Detection of [O III] 88 μm in JADES-GS-z14-0 at z = 14.1793. *Astrophys. J.* **977**, L9 (2024).
13. Ferrara, A., Pallottini, A., & Sommovigo, L. Blue Monsters at z>10: Where has all their dust gone? *Astron. Astrophys.* **694**, 286 (2025)
14. Sanders, R. et al. The AURORA Survey: High-Redshift Empirical Metallicity Calibrations from Electron Temperature Measurements at z=2-10. *arXiv:2508.10099* (2025)
15. Laseter, I. H. et al. JADES NIRSpec Spectroscopy of GN-z11: Lyman-α emission and possible enhanced nitrogen abundance in a z = 10.60 luminous galaxy. *Astron. Astrophys.*, **681**, A70, (2024).
16. Harikane, Y. et al. JWST & ALMA Joint Analysis with [OII] λλ 3726,3729, [OIII] λ4363, [OIII]88 μm, and [OIII]52 μm: Multi-Zone Evolution of Electron Densities at z~0-14 and Its Impact on Metallicity Measurements. arXiv:2505.09186, (2025).
17. Isobe, Y. et al. Redshift Evolution of Electron Density in the Interstellar Medium at z 0-9 Uncovered with JWST/NIRSpec Spectra and Line-spread Function Determinations. *Astrophys. J.*, **956**, 139, (2023).
18. Bunker, A. et al. JADES NIRSpec Spectroscopy of GN-z11: Lyman-α emission and possible enhanced nitrogen

abundance in a z = 10.60 luminous galaxy. *Astron. Astrophys.*, **677**, 88, (2023).

19. Isobe, Y. et al. JADES: nitrogen enhancement in high-redshift broad-line active galactic nuclei. *Mon. Not. R. Astron. Soc.*, **541**, L71 (2025)

20. Cameron, A. et al. Nitrogen enhancements 440 Myr after the big bang: supersolar N/O, a tidal disruption event, or a dense stellar cluster in GN-z11?. *Mon. Not. R. Astron. Soc.*, **523**, 3516 (2023)

21. Le Fèvre, O. *et al.* The ALPINE-ALMA [CII] survey: Multi-wavelength ancillary data and basic physical measurements. *Astron. Astrophys.* **643**, A1 (2020).

22. Béthermin, M. *et al.* The ALPINE-ALMA [CII] survey: Data processing, catalogs, and statistical source properties. *Astron. Astrophys.* **643**, A2 (2020).

23. Faisst, A. L. *et al.* The ALPINE-ALMA [CII] survey: Physical conditions, origins, and fate of the [CII] halos. *Mon. Not. R. Astron. Soc.* **498**, 4192–4210 (2020).

24. Herrera-Camus, R. et al. The ALMA-CRISTAL survey: Gas, dust, and stars in star-forming galaxies when the Universe was ~ 1 Gyr old I. Survey overview and case studies. *Astron. Astrophys,* **699**, A80 (2025).

25. Bouwens, R. J. *et al.* Reionization Era Bright Emission Line Survey: Selection and characterization of luminous interstellar medium reservoirs in the $z > 6.5$ universe. *Astrophys. J.* **931**, 160 (2022).

26. Harikane, Y. et al. Large population of ALMA galaxies at>6 with very high [O III] 88 μm to [C II] 158 μm flux ratios: evidence of extremely high ionization parameter or PDR deficit? *Astrophys. J.* **896**, 93 (2020).

27. Witstok, J. et al. Dual constraints with ALMA: new [O III] 88-μm and dust-continuum observations reveal the ISM conditions of luminous LBGs at z=7. *Mon. Not. R. Astron. Soc.* **515**, 1751–1764 (2022).

28. Inoue, A. K. et al. Detection of an oxygen emission line from a high-redshift galaxy in the reionization epoch. *Science* **352**, 1559–1562 (2016).

29. Algera, H. S. B. et al. Cold dust and low [O III]/[C II] ratios: an evolved star-forming population at redshift 7. *Mon. Not. R. Astron. Soc.* **527**, 6867–6880 (2024).

30. Vallini, L., Ferrara, A., Pallottini, A., Carniani, S., & Gallerani, S. High [O III]/[C II] surface brightness ratios trace early starburst galaxies. *Mon. Not. R. Astron. Soc.* **505**, 5543–5553 (2021).

31. Katz, H. et al. The nature of high [OIII]88 μm/[CII]158 μm galaxies in the epoch of reionization: Low carbon abundance and a top-heavy IMF?. *Mon. Not. R. Astron. Soc.,* **510**, 5603 (2022)

32. Gelli, V., Mason, C., & Hayward, C. C. The Impact of Mass-dependent Stochasticity at Cosmic Dawn. *Astrophys. J.* **975**, 192 (2024).

33. Sommovigo, L. et al. Dust temperature in ALMA [C II]-detected high-z galaxies. *Mon. Not. R. Astron. Soc.* **503**, 4878–4889 (2021).

34. Li, J. et al. The ALMA-CRISTAL Survey: Spatially Resolved Star Formation Activity and Dust Content in 4 < z < 6 Star-forming Galaxies. *Astrophys. J.* **976**, 70 (2024).

35. Lines, N. E. P. et al. JWST PRIMER: A lack of outshining in four normal z = 4–6 galaxies from the ALMA-CRISTAL Survey. *Mon. Not. R. Astron. Soc.* **539**, 2685–2706 (2025).

36. Villanueva, V. et al. The ALMA-CRISTAL survey: Dust temperature and physical conditions of the interstellar medium in a typical galaxy at z=5.66. *Astron. Astrophys.* **691**, A133 (2024).

37. Mitsuhashi, I. et al. SERENADE. II. An ALMA Multiband Dust Continuum Analysis of 28 Galaxies at 5 < z < 8 and the Physical Origin of the Dust Temperature Evolution. *Astrophys. J.* **971**, 161 (2024).

38. Pallottini, A. et al. A survey of high-z galaxies: SERRA simulations. *Mon. Not. R. Astron. Soc.* **513**, 5621–5641 (2022).

39. Vallini, L. Spatially resolved [CII]–gas conversion factor in early galaxies. *Astron. Astrophys.*, **700**, A117, (2025).

40. Nakazato, Y., Yoshida, N., & Ceverino, D. Simulations of High-redshift [O III] Emitters: Chemical Evolution and Multiline Diagnostics. *Astrophys. J.* **953**, 140 (2023).

41. Katz, H. et al. First insights into the ISM at z>8 with JWST: possible physical implications of a high [O III] λ4363/[O III] λ5007. *Mon. Not. R. Astron. Soc.* **518**, 592–604 (2023).

42. Schimek, A. et al. Constraining the physical properties of gas in high-z galaxies with far-infrared and submillimetre line ratios. *Astron. Astrophys.* **687**, L10 (2024).

43. Tadaki, K.-i. et al. CNO Emission of an Unlensed Submillimeter Galaxy at z=4.3. *Astrophys. J.* **876**, 1 (2019).

44. Vallini, L., Ferrara, A., Pallottini, A., Carniani, S., & Gallerani, S. Star formation law in the epoch of reionization from [C II] and C III] lines. *Mon. Not. R. Astron. Soc.* **495**, L22–L26 (2020).

45. Jones, T. et al. The Mass–Metallicity Relation at z ≈ 8: Direct-method Metallicity Constraints and Near-future Prospects. *Astrophys. J.*, **903**, 150, (2020).

46. Aravena, M. et al. The ALMA Reionization Era Bright Emission Line Survey: The molecular gas content of galaxies at z=7. *Astron. Astrophys.* **682**, A24 (2024).

47. Dessauges-Zavadsky, M. et al. The ALPINE-ALMA [C II] survey: Molecular gas budget in the early Universe as traced by [C II]. *Astron. Astrophys.* **643**, A5 (2020).

48. Herrera-Camus, R. et al. Kiloparsec view of a typical star-forming galaxy when the Universe was ~1 Gyr old. I. Properties of outflow, halo, and interstellar medium. *Astron. Astrophys.* **649**, A31 (2021).

49. Heintz, K. E. et al. Measuring the HI Content of Individual Galaxies Out to the Epoch of Reionization with [C II]. *Astrophys. J.* **922**, 147 (2021).

50. Wilson, S. A high-redshift calibration of the [O I]-to-H I conversion factor in star-forming galaxies *Astron. Astrophys.* **685**, A30 (2024).

51. Heintz, K. E. et al. Strong damped Lyman-α absorption in young star-forming galaxies at redshifts 9 to 11. *Science*, **384**, 6698, (2024)

52. Carpenter, J. et al. ALMA Wideband Sensitivity Upgrade: Expanding ALMA's Capability for Cosmic Discovery. In Physics and Chemistry of Star Formation: The Dynamical ISM Across Time and Spatial Scales, Ossenkopf-Okada, V et al. (eds), p304. Cologne: Universitäts- und Stadtbibliothek Köln (2023).

53. Rizzo, F. et al. Dynamical properties of z ~ 4.5 dusty star-forming galaxies and their connection with local early-type galaxies. *Mon. Not. R. Astron. Soc.* **507**, 3952–3984 (2021).

54. Neeleman, M., et al. A cold, massive, rotating disk galaxy 1.5 billion years after the Big Bang. *Nature* **581**, 269–272 (2020).

55. Danhaive, A. L. et al. The dawn of disks: unveiling the turbulent ionised gas kinematics of the galaxy population at z~4–6 with JWST/NIRCam grism spectroscopy. arXiv:2503.21863, (2025)

56. Kohandel, M. et al. Dynamically cold disks in the early Universe: Myth or reality? *Astron. Astrophys.* **685**, A72 (2024).

57. Lee, L. et al. The ALMA-CRISTAL survey: Resolved kinematic studies of main sequence star-forming galaxies at 4< z<6. *Astron. Astrophys.* **701**, 260, (2025)

58. Ferreira, L., et al. The JWST Hubble Sequence: The Rest-frame Optical Evolution of Galaxy Structure at 1.5 < z < 6.5. *Astrophys. J.* **955**, 94 (2023)

59. Fujimoto, S., et al. High-resolution observations of a strongly lensed, low-luminosity galaxy at z=6.072: Insights into star-forming clumps and galaxy dynamics. *Nat. Astron.* In press (2025).

60. Rowland, L. E., et al. The gas dynamics of REBELS-25, a [C II]-luminous galaxy at z=7.3065. *Mon. Not. R. Astron. Soc.*, **535**, 2068 (2024).

61. Förster Schreiber, N. M., Wuyts, S. Star-Forming Galaxies at Cosmic Noon. *Annu. Rev. Astron. Astrophys.*, **58**, 661-705 (2020).

62. Übler, H., et al. GA-NIFS: NIRSpec reveals evidence for non-circular motions and AGN feedback in GN20. *Mon. Not. R. Astron. Soc.*, **533**, 4287-4301 (2024).

63. Genzel, R. et al. Evidence for Large-scale, Rapid Gas Inflows in z ~ 2 Star-forming Disks. *Astrophys. J.* **957**, 48 (2023).

64. Fujimoto, S., et al. The ALPINE-ALMA [C II] Survey: Size of Individual Star-forming Galaxies at z = 4-6 and Their Extended Halo Structure. *Astrophys. J.*, **900**, 1 (2020).

65. Lambert, T. S. et al. An extended [C II] halo around a massive star-forming galaxy at z=5.3. *Mon. Not. R. Astron. Soc.* **518**, 3183–3191 (2024).

66. Ikeda, R., et al. The ALMA-CRISTAL Survey: Spatial extent of [CII] line emission in star-forming galaxies at z = 4-6. *Astron. Astrophys.*, **693**, A237 (2025).

67. Pizzati, E., et al. [C II] Haloes in ALPINE galaxies: smoking-gun of galactic outflows?. *Astron. Astrophys.*, **673**, A39 (2023).

68. Rey, M., et al. (2024). *ARCHITECTS I: impact of subgrid physics on the simulated properties of the circumgalactic medium. Mon. Not. R. Astron. Soc.* **543**, 12-27 (2025).

69. Solimano, M., et al. The ALMA-CRISTAL survey: Discovery of a 15 kpc-long gas plume in a z = 4.54 Lyman-α blob. *Astron. Astrophys.* **689**, A145 (2024).

70. Solimano, M., et al. A hidden active galactic nucleus powering bright [O III] nebulae in a protocluster at z = 4.5 revealed by JWST. *Astron. Astrophys.* **693**, A70 (2025).

71. Harikane, Y., et al. A JWST/NIRSpec First Census of Broad-line AGNs at z = 4-7: Detection of 10 Faint AGNs with M BH 106-108 M ⊙ and Their Host Galaxy Properties. *Astrophys. J.* **959**, 39 (2023).

72. Maiolino, R., et al. JADES: The diverse population of infant black holes at 4 < z < 11: Merging, tiny, poor, but mighty. *Astron. Astrophys.* **691**, A145 (2024).

73. Reines, A. E., & Volonteri, M. Relations between Central Black Hole Mass and Total Galaxy Stellar Mass in the Local Universe. *Astrophys. J.* **813**, 82 (2015).

74. Li, J., et al. Tip of the Iceberg: Overmassive Black Holes at 4 < z < 7 Found by JWST Are Not Inconsistent with the Local Relation. *Astrophys. J. 981*, 19, (2025).

75. Volonteri, M., et al. What if young z > 9 JWST galaxies hosted massive black holes?. *Mon. Not. R. Astron. Soc.* **521**, 241 (2023).

76. Lauer, T. R. et al. The Masses of Nuclear Black Holes in Luminous Elliptical Galaxies and Implications for the Space Density of the Most Massive Black Holes. *Astrophys. J.* **662**, 808–834 (2007).

77. Dayal, P., et al. Exploring a primordial solution for early black holes detected with JWST. *Astron. Astrophys.* **690**, A182 (2024).

78. Natarajan, P., et al. First light of supermassive black holes: Evidence for a heavy-seed origin. *Astrophys. J.* **960**, L1 (2024).

79. Labbé, I., et al. A population of red candidate massive galaxies 600 Myr after the Big Bang. *Nature*, **616**, 266–269 (2023).

80. Matthee, J., et al. Little Red Dots: An Abundant Population of Faint Active Galactic Nuclei at z ~ 5 Revealed by the EIGER and FRESCO JWST Surveys. *Astrophys. J.*, **963**, 129 (2024).

81. Setton, D. et al. A confirmed deficit of hot and cold dust emission in the most luminous Little Red Dots. *Astrophys. J. Letters*, **991**, L10 (2025).

82. Inayoshi, K. & Maiolino, R. Extremely Dense Gas around Little Red Dots and High-redshift Active Galactic Nuclei: A Nonstellar Origin of the Balmer Break and Absorption Features. *Astrophys. J. Letters*, **980**, L27, (2025).

83. Akins, H. B., et al. COSMOS-Web: The over-abundance and physical nature of 'little red dots'—Implications for early galaxy and SMBH assembly. Astrophys.l J., **991**, 1 (2025).

84. Pérez-González, P. G., et al. What Is the Nature of Little Red Dots and What Is Not, MIRI SMILES Edition. *Astrophys. J.*, **968** (2024)

85. Juodžbalis, I. et al. A direct black hole mass measurement in a Little Red Dot at the Epoch of Reionization. arXiv:2508.21748, (2025).

86. Lupi, A., et al. Size matters: are we witnessing super-Eddington accretion in high-redshift black holes from JWST?. *Astron. Astrophys*, **689**, A128. (2024)

87. Greene, J. E., & Ho, L. C. The Mass Function of Active Black Holes in the Local Universe. *Astrophys. J.*, **667**(1), 131. (2007).

88. Buchner, J. et al. Genuine Retrieval of the AGN Host Stellar Population (GRAHSP). *Astron. Astrophys.*, **692**, A161, (2025).

89. Übler, H. et al. GA-NIFS: JWST discovers an offset AGN 740 million years after the Big Bang. *Mon. Not. R. Astron. Soc.*, **531**, 355–365. (2024).

90. Abuter, R. et al. A dynamical measure of the black hole mass in a quasar 11 billion years ago. *Nature*, **627**, 281–285. (2024).

91. Greene, J. E., Strader, J., & Ho, L. C. Intermediate-Mass Black Holes. *Annu. Rev. Astron. Astrophys.*, **58**, 257–312 (2020).

92. Carniani, S. et al. JADES: The incidence rate and properties of galactic outflows in low-mass galaxies across z<9. *Astron. Astrophys*, **685**, A99 (2024).

93. Fluetsch, A. et al. Cold molecular outflows in the local Universe and their feedback effect on galaxies. *Mon. Not. R. Astron. Soc.*, **483**(4), 4586–4614 (2019).

94. Spilker, J. S., et al. Ubiquitous Molecular Outflows in z > 4 Massive, Dusty Galaxies. II. Momentum-driven Winds Powered by Star Formation in the Early Universe. *Astrophys. J.*, **905**, 86, (2020).

95. Ginolfi, M. et al. The ALPINE-ALMA [C II] survey: Star-formation-driven outflows and circumgalactic enrichment in the early Universe. *Astron. Astrophys.*, **633**, A90, (2020).

96. Birkin, J. et al. The ALMA-CRISTAL survey: weak evidence for star-formation driven outflows in z ~5 main-sequence galaxies. *Astrophys. J.* **985**, 243(2025).

97. Parlanti, E. et al. GA-NIFS: Multi-phase analysis of a star-forming galaxy at z~5.5. Accepted in *Astron. Astrophys.* **696**, 6, (2025).

98. Faisst, A. et al. The ALPINE-CRISTAL-JWST Survey: JWST/IFU Optical Observations for 18 Main-Sequence Galaxies at z=4-6. *arXiv:2510.16111* (2025)

99. Fujimoto, S. et al. The ALPINE-CRISTAL-JWST Survey: NIRSpec IFU Data Processing and Spatially-resolved Views of Chemical Enrichment in Normal Galaxies at z=4-6. arXiv:2510.16116 (2025)

100. Boquien, M. et al. Python Code Investigating GALaxy Emission (CIGALE). *Astron. Astrophys.* **622**, A103 (2019).